# Coupling parameter lower bound in Yukawa-Vlasov plasmas

Kenneth I. Golden
Department of Mathematics and Statistics and Department of Physics
University of Vermont, Burlington, Vermont 05401-1455, USA

and

Gabor J. Kalman
Department of Physics
Boston College, Chestnut Hill, Massachusetts 02467, USA

ABSTRACT

We have analyzed the Vlasov dispersion relation for Yukawa plasmas in three and two dimensions primarily for the purpose of identifying coupling parameter domains where the existence of well-developed collective excitations is forbidden or allowed. First, we have established a rigorous lower bound for the coupling parameter, below which there can be no real solution to the dispersion relation. In the coupling domain where weakly damped solutions do exist, we have focused on the long-wavelength acoustic regime where we have established more restrictive lower-bound estimates of the coupling parameter. We have also derived a general formula for the corresponding acoustic phase velocity, valid over a wide range of coupling parameter/screening parameter ratios above the more restrictive lower bound.

In this paper, we analyze the Vlasov dispersion relation for Yukawa plasmas [1] primarily for the purpose of identifying coupling parameter-screening parameter domains where the existence of a viable longitudinal collective mode is forbidden or allowed.

In complex plasmas characterized by the Yukawa potential

$$\varphi(r) = Q^2 \frac{\exp(-\kappa r)}{r} \tag{1}$$

($Q$ is the charge of a dust grain and $\kappa$ is the screening parameter), the collective mode behavior in the low coupling regime is customarily calculated from the zeros of the Vlasov dielectric response function [2], viz.,

$$\varepsilon(\mathbf{k},\omega) = 1 - \varphi(k)\chi(\mathbf{k},\omega) = 0, \tag{2}$$

$$\chi(\mathbf{k},\omega) = -\frac{1}{m}\int d\mathbf{v}\, \frac{\mathbf{k}\cdot\partial F(v)/\partial \mathbf{v}}{\omega - \mathbf{k}\cdot\mathbf{v}}; \tag{3}$$

$\varphi(k)$ is the Fourier transform of (1), $\chi(\mathbf{k},\omega)$ is the screened (total) density respone function, and $F(v)$ is the Maxwellian distribution function normalized to the average particle density $n$. In this paper, we will analyze the zeros of Eq. (2) in the long-wavelength acoustic domain from the perspective of establishing bounds on the coupling parameter invoked by the Vlasov approximation.

On the premise that the Landau damping is weak, seeking the zeros of (2) amounts to solving

$$\frac{1}{\varphi(k)} = \operatorname{Re}\chi(k,\omega(k)),, \tag{4}$$

for the (real) oscillation frequency $\omega(k)$, accompanied by the calculation of the companion damping rate

$$\gamma(k) = -\frac{\operatorname{Im} \varepsilon(k,\omega(k))}{(\partial/\partial\omega)\operatorname{Re}\varepsilon(k,\omega)\big|_{\omega=\omega(k)}} = -\frac{\operatorname{Im}\chi(k,\omega(k))}{(\partial/\partial\omega)\operatorname{Re}\chi(k,\omega)\big|_{\omega=\omega(k)}} < 0. \quad (5)$$

In deriving Eqs. (4) and (5) we have invoked the well-known weak-damping hypothesis

$$\left|\operatorname{Im}\chi(k,\omega(k))\right| << \left|\operatorname{Re}\chi(k,\omega(k))\right|, \qquad \left|\gamma(k)\right| << \omega(k) \quad (6)$$

We consider both three- and two-dimensional Yukawa plasmas characterized by the Fourier-transformed interaction potentials

$$\varphi(k) = \frac{4\pi Q^2}{\kappa^2 + k^2} \qquad \text{3D} \quad (7)$$

$$\varphi(k) = \frac{2\pi Q^2}{\sqrt{\kappa^2 + k^2}} \qquad \text{2D} \quad (8)$$

The following dispersion relations then result from Eqs. (7) - (9):

$$\frac{\bar{\kappa}^2 + \bar{k}^2}{3\Gamma} = \frac{1}{\beta n}\operatorname{Re}\chi(x) \qquad \text{3D} \quad (9)$$

$$\frac{\sqrt{\bar{\kappa}^2 + \bar{k}^2}}{2\Gamma} = \frac{1}{\beta n}\operatorname{Re}\chi(x); \qquad \text{2D} \quad (10)$$

$$\frac{1}{\beta n}\operatorname{Re}\chi(x) = -1 + x\frac{1}{\sqrt{\pi}} P\int_{-\infty}^{\infty} dt \frac{\exp(-t^2)}{x-t}, \quad (11)$$

$$\frac{1}{\beta n}\operatorname{Im}\chi(x) = -\sqrt{\pi}x\exp(-x^2); \quad (12)$$

$x = (\omega/k)\sqrt{\beta m/2}$, $\bar{\kappa} = \kappa a$, $\bar{k} = ka$, $(4/3)\pi n a^3 = 1$ and $\pi n a^2 = 1$ define the Wigner-Seitz radius $a$ in three and two dimensions, respectively; $\Gamma = \beta Q^2/a$ is the conventional coupling parameter with $\beta = 1/k_B T$. The function $(1/\beta n)\operatorname{Re}\chi(x)$, shown plotted in Fig. 1, reaches its maximum value $(1/\beta n)\operatorname{Re}\chi(x)\big|_{\max} = 0.2847$ at

$x = 1.5$.

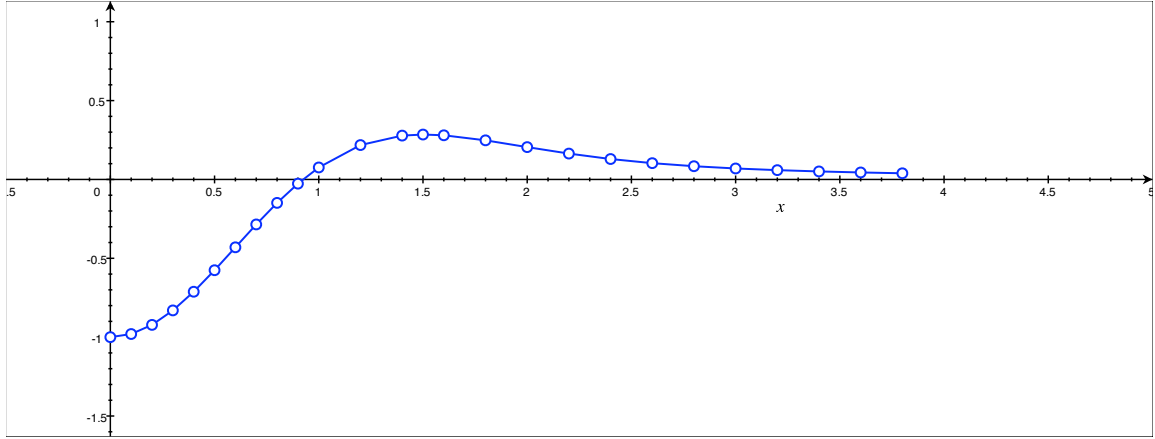

Figure 1: $(1/\beta n)\operatorname{Re}\chi(x)$ as a function of $x = (\omega/k)\sqrt{\beta m/2}$.

Now, for $\bar{k}$ fixed, $x$ varies with $\omega$ only. Consequently, for a given assigned pair of values to $(\Gamma, \bar{\kappa})$, the left-hand-side term of dispersion relation (9) [or (10) in 2D], being independent of $\omega$, would graph as a horizontal line above the $x$ axis in Fig. 1. Then clearly there can be no solution to Eq. (9) [or (10) in 2D] if the horizontal line is situated above $(1/\beta n)\operatorname{Re}\chi(x)|_{\max} = 0.2847$, that is, if

$$\frac{\bar{k}^2 + \bar{\kappa}^2}{3\Gamma} \geq \frac{\bar{\kappa}^2}{3\Gamma} > 0.2847 \qquad \underline{3D} \qquad (13)$$

$$\frac{\sqrt{\bar{k}^2 + \bar{\kappa}^2}}{2\Gamma} \geq \frac{\bar{\kappa}}{2\Gamma} > 0.2847 \qquad \underline{2D} \qquad (14)$$

Thus, for $\bar{k} > 0$, the conditions

$$\Gamma < 1.17\bar{\kappa}^2 \qquad \underline{3D} \qquad (15)$$

$$\Gamma < 1.756\bar{\kappa} \qquad \underline{2D} \qquad (16)$$

are *sufficient* to guarantee that there can be no solutions to Eqs. (9) and (10). In the long-wavelength $(k \to 0)$ limit, which is especially of interest in the present work, conditions (15) and (16) become *necessary* as well.

Concentrating on the long-wavelength domain where the dispersion is acoustic, i.e., $\omega \propto k$, we now consider the possibilities when the inequalities (15) and (16) are not satisfied. While

$$\Gamma = 1.17\bar{\kappa}^2 \qquad \underline{3D} \qquad (17)$$

$$\Gamma = 1.756\bar{\kappa} \qquad \underline{2D} \qquad (18)$$

do indeed satisfy dispersion relations (9) and (10), they are obviously ruled out as possibilities since the derivative $(\partial/\partial\omega)\mathrm{Re}\,\chi(k,\omega)$ can never be positive definite at $\omega = \omega(k)$ [see Eqs. (5), (6), (12)]. Consequently, only the descending portion of the curve in Fig. 1 is relevant, and a more restricted lower bound on the coupling parameter can be estimated by selecting the smallest value of *x* on the descending portion that will ensure that the Landau damping is sufficiently weak to allow the formation of a viable acoustic mode. The following Table facilitates that selection.

| $x$ | $\frac{1}{\beta n}\mathrm{Re}\,\chi(x)$ | $\frac{1}{\beta n}\mathrm{Im}\,\chi(x)$ | $\frac{1}{\beta n}\frac{\partial}{\partial x}\mathrm{Re}\,\chi(x)$ | $\frac{\gamma}{\omega}$ |
|---|---|---|---|---|
| 1.6 | 0.280 | -0.2193 | -0.0960 | -1.4280 |
| 1.8 | 0.248 | -0.1250 | -0.1994 | -0.3483 |
| 2.0 | 0.205 | -0.0649 | -0.2175 | -0.1492 |
| 2.2 | 0.164 | -0.0308 | -0.1926 | -0.0728 |
| 2.4 | 0.130 | -0.0134 | -0.1509 | -0.0372 |
| 2.6 | 0.103 | -0.0053 | -0.1128 | -0.0181 |
| 2.8 | 0.084 | -0.0020 | -0.0842 | -0.0078 |
| 3.0 | 0.070 | -0.0007 | -0.0611 | -0.0036 |
| 3.2 | 0.059 | -0.0002 | -0.0467 | -0.0014 |
| 3.4 | 0.051 | -0.0001 | -0.0344 | -0.0005 |
| 3.6 | 0.044 | $-1.5\times10^{-5}$ | -0.0299 | -0.0001 |
| 3.8 | 0.039 | $-3.6\times10^{-6}$ | -0.0237 | $-4\times10^{-5}$ |

We observe that the Landau damping diminishes, as it should, with increasing *x*. To be on the safe side, we select $x = 2.2$ as the value beyond which the acoustic mode can be considered to be viable. On this basis, we can state that the realization of a long-lived acoustic mode requires that

$$\frac{\bar{\kappa}^2}{3\Gamma} < 0.164, \qquad \underline{\text{3D}} \qquad (19)$$

$$\frac{\bar{\kappa}}{2\Gamma} < 0.164, \qquad \underline{\text{2D}} \qquad (20)$$

resulting in the more restrictive lower bound estimates:

$$\Gamma > 2.03\bar{\kappa}^2 \qquad \underline{\text{3D}} \qquad (21)$$

$$\Gamma > 3.05\bar{\kappa} \qquad \underline{\text{2D}} \qquad (22)$$

The subsequent calculation of the acoustic phase velocity, valid for $x \geq 2.4$, is carried out in two stages: First, we solve the dispersion relation (for *x*) resulting from the combination of Eq. (9) [or (10)] in 2D) and the large-*x* expansion of (11):

$$\frac{1}{\beta n}\operatorname{Re}\chi(x) \approx \frac{1}{2x^2} + \frac{3}{4x^4} + \cdots \qquad (23)$$

Next, we introduce into the solution a higher-order correction proportional to $\bar{\kappa}^4/\Gamma^2$ [or $\bar{\kappa}^2/\Gamma^2$ in 2D] to guarantee agreement with the *x* entries in the Table. The following sound speed formulas (with accuracy exceeding 99%) then result:

$$\beta m s^2 = \frac{3\Gamma}{2\bar{\kappa}^2}\left[1 + \sqrt{1 + 4\frac{\bar{\kappa}^2}{\Gamma} + 9\frac{\bar{\kappa}^4}{\Gamma^2}}\right] \qquad \underline{\text{3D}} \qquad (24)$$

$$\beta m s^2 = \frac{\Gamma}{\bar{\kappa}}\left[1 + \sqrt{1 + 6\frac{\bar{\kappa}}{\Gamma} + 21\frac{\bar{\kappa}^2}{\Gamma^2}}\right] \qquad \underline{\text{2D}} \qquad (25)$$

**Acknowledgements**

This work has been partially supported by NSF Grants No. PHY-0812956, No. PHY-0813153, and No. PHY-1105005.